\documentclass[journal]{IEEEtran}
\usepackage{amsmath, amssymb, amsthm, mathtools}
\usepackage{graphicx, subfigure, xcolor}
\usepackage{relsize, paralist, hyperref, cite}

\newtheorem{theorem}{Theorem}

\newtheorem{proposition}[theorem]{Proposition}

\newtheorem{remark}{Remark}

\newcommand{ \C }{ \mathcal C }
\newcommand{ \B }{ \mathcal B }
\newcommand{ \Bb }{ \tilde{\mathcal B} }
\newcommand{ \A }{ \mathcal A }
\newcommand{ \Sp }{ \mathcal S }
\newcommand{ \myx }{ \boldsymbol{x} }
\newcommand{ \myxt }{ \boldsymbol{\tilde{x}} }
\newcommand{ \xt }{ \tilde{x} }
\newcommand{ \myy }{ \boldsymbol{y} }
\newcommand{ \myyt }{ \boldsymbol{\tilde{y}} }
\newcommand{ \prob }{ \textnormal{Pr} }
\newcommand{ \bs }[1]{ \boldsymbol{#1} }
\newcommand{ \myqed }{ \hfill $\blacktriangle$ }
\newcommand{ \defeq }{ \coloneqq }

\begin{document}

\title{Zero-Error Capacity of Duplication Channels}

\author{Mladen~Kova\v{c}evi\'c
				
\thanks{Date: July 22, 2019.}
\thanks{The author is with the BioSense Institute, University of Novi Sad, 21000 Novi Sad, Serbia
				(email: kmladen@uns.ac.rs).}
}


\maketitle

\begin{abstract}
\boldmath{%
This paper is concerned with the problem of error-free communication over
the i.i.d. duplication channel which acts on a transmitted sequence
$ x_1 \cdots x_n $ by inserting a random number of copies of each symbol
$ x_i $ next to the original symbol.
The random variables representing the numbers of inserted copies at each
position $ i $ are independent and take values in $ \{0, 1, \ldots, r\} $,
where $ r $ is a fixed parameter.
A more general model in which blocks of $ \ell $ consecutive symbols are
being duplicated, and which is inspired by DNA-based data storage systems
wherein the stored molecules are subject to tandem-duplication mutations,
is also analyzed.
A construction of optimal codes correcting \emph{all} patterns of errors
of this type is described, and the zero-error capacity of the duplication
channel---the largest rate at which information can be transmitted through
it in an error-free manner---is determined for each $ \ell $ and $ r $.%
}
\end{abstract}


\section{Introduction and Preliminaries}
\label{sec:intro}

\IEEEPARstart{C}{hannels} with synchronization errors are a class of communication
models that attempt to capture the impairments in the received signal caused
by the loss of synchronization at the symbol level and similar phenomena.
Starting with Levenshtein's work on the insertion/deletion channel \cite{levenshtein2},
this and related models have inspired a significant amount of research due to
both their practical relevance and theoretical appeal (see, e.g., the surveys
\cite{mercier_survey, mitzenmacher_survey}).
However, their analysis is in general complicated and they have become somewhat
notorious for resisting the attempts of researchers to quantify even their basic
information-theoretic limits.

The present paper is focused on a particular type of channels with synchronization
errors that are usually referred to as duplications, repetitions, or sticky insertions.
These channels model the communication scenario in which the receiver is sampling
the received (baseband) signal at a rate faster than the transmitter's clock,
causing some of the symbols to be read more than once and thus resulting in a
longer string in which some of the transmitted symbols are erroneously repeated
several times.
For the purpose of generality, we shall in fact consider a model in which \emph{blocks}
of $ \ell $ consecutive symbols are being duplicated, for any fixed parameter
$ \ell \geq 1 $.
These kinds of impairments have also attracted some interest lately because they
correspond to tandem-duplication mutations of DNA sequences \cite{mundy} and are
thus of potential relevance for DNA-based data storage applications
\cite{farnoud,jain,jain2,kovacevic+tan2}.

With the above-mentioned information transmission and storage scenarios as
motivating examples, we study the problem of reliable communication in the
presence of duplication errors.
Our results include a construction of optimal codes correcting \emph{all} possible
patterns of duplication errors, and a consequent characterization of the zero-error
capacity of duplication channels, which is the largest rate at which information
can be transmitted through them in an error-free manner.
The methods used in the analysis are direct extensions of the well-known methods
for discrete memoryless channels (DMCs) \cite{shannon}.
Nonetheless, the obtained results are interesting in that they provide
\emph{one of the first examples of a realistic and non-trivial discrete-time
channel with synchronization errors for which such a \emph{coding theorem} has
been established}.
This is also one of the rare cases where zero-error capacity is easier to
determine than the Shannon (vanishing-error) capacity.

In the remainder of this section we describe the channel model, introduce
basic terminology, and list the relevant literature.
Our main results concerning zero-error codes and capacity of duplication
channels are presented in Section~\ref{sec:main}, while Section~\ref{sec:cw}
presents a refinement of these results pertaining to constant-weight codes.

\subsection{The Channel Model}
\label{sec:channel}

The channel alphabet is denoted by $ \A_q \defeq \{0, 1, \ldots, q-1\} $, and
the set of all words over $ \A_q $ by $ \A_q^* \defeq \bigcup_{i=0}^\infty \A_q^i $.

We start with the description of the model in the special case $ \ell = 1 $.
Consider the channel that acts on an input string $ \myxt = \xt_1 \cdots \xt_n $ by
inserting a random number of copies of each symbol $ \xt_i $ next to the original
symbol.
More precisely, for every $ i = 1, \ldots, n $, the channel inserts $ c_i $ copies
of the symbol $ \xt_i $ next to it, where it is assumed that $ \prob\{c_i=k\} > 0 $
for $ k = 0, 1, \ldots, r $ and $ \sum_{k=0}^r \prob\{c_i=k\} = 1 $, and that the
random variables $ c_i $ and $ c_j $ are independent for any $ i \neq j $.
Thus, the maximum number of inserted copies at each position is $ r \in \{1, 2, \ldots\} $,
a fixed parameter%
\footnote{The assumption that the number of inserted copies of each symbol is
bounded by a parameter $ r < \infty $ is realistic in sticky-insertion channels
where such errors occur due to the mismatch between the transmitter's and the
receiver's clock.
Namely, it is reasonable to expect that this mismatch is small and that only a
bounded number of copies of a symbol may be inserted at each position.
In fact, the case $ r = 1 $, where each input symbol is either duplicated (once)
or is left intact, is likely to apply in practice.
}.
The specific values of the probabilities $ \prob\{c_i=k\} $ are irrelevant for
the problems addressed in this paper, only the assumption that they are all
strictly positive for $ k = 0, 1, \ldots, r $ is needed.
We refer to the channel just described as the $ q $-ary $ (1, r) $-duplication
channel, where the word $ q $-ary is often omitted as the alphabet is understood
from the context.

As an example, consider the following input string $ \myxt \in \A_3^{7} $ and
the corresponding string $ \myyt \in \A_3^{11} $ obtained at the output of the
ternary $ (1, 2) $-duplication channel:
\begin{align}
\nonumber
  \myxt  \ &= \  1 \ 0 \ 1 \ 2 \ 2 \ 1 \ 2   \\
\label{eq:xyt}
  \myyt  \ &= \  1 \ 0 \ \underline{0} \ 1 \ \underline{1} \ \underline{1} \ 2 \ 2 \ \underline{2} \ 1 \ 2 .
\end{align}
The inserted duplicates are underlined;
the total number of duplication errors that occurred in the channel is $ 1 + 2 + 1 = 4 $.

By using the transformation
$ \phi : \A_q^* \to \A_q^* $, $ \myxt \mapsto \myx $,
defined by $ x_i = \tilde{x}_i - \tilde{x}_{i-1} $,
$ 1 \leq i \leq n $, where subtraction is performed modulo $ q $ and it is
understood that $ \tilde{x}_0 = 0 $, it is easy to see that duplication
errors are essentially equivalent to insertions of zeros \cite{dolecek+anantharam}.
For example, for the strings in \eqref{eq:xyt} we have:
\begin{align}
\nonumber
  \myx  \ &= \  1 \ 2 \ 1 \ 1 \ 0 \ 2 \ 1   \\
\label{eq:xy}
  \myy  \ &= \  1 \ 2 \ \underline{0} \ 1 \ \underline{0} \ \underline{0} \ 1 \ 0 \ \underline{0} \ 2 \ 1 .
\end{align}
We can thus define the $ q $-ary $ (1, r) $-$ 0 $-insertion channel as the
channel which acts on an input string $ \myx = x_1 \cdots x_n $ by inserting
$ c_i $ zeros after the symbol $ x_i $ (independently of everything else),
where $ \prob\{c_i=k\} > 0 $ for $ k = 0, 1, \ldots, r $ and
$ \sum_{k=0}^r \prob\{c_i=k\} = 1 $.
By using the bijective mapping $ \phi $ one can switch between the
$ (1, r) $-duplication and the $ (1, r) $-$ 0 $-insertion channels, as well
as between the corresponding codes, so we shall use both of these equivalent
descriptions interchangeably.

Consider a string $ \myx = \sigma_1\,0^{u_1} \,\cdots\, \sigma_w\,0^{u_w} $,
where $ \sigma_i \in \A_q\!\setminus\!\{0\} $ and $ 0^u $ denotes a block
of $ u $ zeros.
This representation is particularly useful for the $ 0 $-insertion channel
as this channel affects only the runs of zeros in an input string $ \myx $.
A simple but important observation here is that the effect of the channel
on the segment $ \sigma_i\,0^{u_i} $ is independent of its effect on any
other segment $ \sigma_j\,0^{u_j} $, $ i \neq j $.
Therefore, the $ (1, r) $-$ 0 $-insertion channel (with an additional
assumption that the first symbol of every codeword is non-zero) is equivalent
to a DMC with alphabet
$ \{ \sigma\,0^u : \sigma \in \A_q\!\setminus\!\{0\}, u \geq 0 \} $.
Apart from the alphabet of this DMC being infinite, each element of the
alphabet has a ``cost'' assigned to it \cite{mitzenmacher}.
Namely, the element $ \sigma\,0^u $ is assigned a cost of $ u+1 $, equal
to its length when considered as a string over $ \A_q $, i.e., the cost is
the number of symbols from $ \A_q $ that need to be transmitted in order
for this element to be ``transmitted'' over the equivalent DMC.

Let us now introduce a more general model in which blocks of length $ \ell $
are being duplicated, where $ \ell \in \{1, 2, \ldots\} $ is a fixed parameter.
This generalization is primarily of theoretical value but, as mentioned in the
introductory part of the paper, it is of potential interest in DNA-based data
storage systems wherein the stored molecules are subject to tandem-duplication
mutations \cite{jain2}.
Define the $ (\ell, r) $-duplication channel as the channel that acts on an
input string $ \myxt $ by inserting after the symbol $ \xt_i $ at most $ r $
copies of the substring $ \xt_{i-\ell+1} \cdots \xt_i $, for every
$ i = \ell, \ldots, n $.
For example, consider the following input string $ \myxt \in \A_3^{7} $ and
the corresponding string $ \myyt \in \A_3^{16} $ obtained at the output of the
ternary $ (3, 2) $-duplication channel:
\begin{align}
\nonumber
  \myxt  \ &= \  1 \ 0 \ 1 \ 2 \ 2 \ 1 \ 2   \\
\label{eq:xytl}
  \myyt  \ &= \  1 \ 0 \ 1 \ 2 \ \underline{0 \ 1 \ 2} \ 2 \ 1 \ \underline{2 \ 2 \ 1} \ \underline{2 \ 2 \ 1} \ 2 .
\end{align}
The inserted duplicates of blocks of length $ \ell = 3 $ are underlined;
the total number of (tandem) duplication errors that occurred in the channel is
$ 1 + 2 = 3 $.
Again, by using a suitable transformation $ x_i = \phi_\ell(\xt_i) = \xt_i - \xt_{i-\ell} $,
one can show that this channel is equivalent to the channel which inserts
up to $ r $ blocks $ 0^\ell $ after the symbol $ x_i $, for every
$ i = \ell, \ldots, n $ \cite{jain2}.
For example, for the strings in \eqref{eq:xyt} we have:
\begin{align}
\nonumber
  \myx  \ &= \  1 \ 0 \ 1 \ 1 \ 2 \ 0 \ 0   \\
\label{eq:xyl}
  \myy  \ &= \  1 \ 0 \ 1 \ 1 \ \underline{0 \ 0 \ 0} \ 2 \ 0 \ \underline{0 \ 0 \ 0} \ \underline{0 \ 0 \ 0} \ 0 .
\end{align}
For simplicity, we shall consider a slight variation of the latter model:
let $ (\ell, r) $-$ 0 $-insertion channel be the channel which inserts up to
$ r $ blocks $ 0^\ell $ after the symbol $ x_i $, for every $ i = 1, \ldots, n $.
The difference is only in the possibility to insert blocks $ 0^\ell $ after
the first $ \ell - 1 $ symbols, which is irrelevant for the asymptotic analysis
(see Remark \ref{rem:assmp} ahead on how this assumption can be dropped).

\subsection{Terminology and Conventions}
\label{sec:terminology}

Two strings $ \myx, \myy \in \A_q^* $ are said to be confusable in a given
communication channel if they can produce the same string at the output of
that channel.
They are said to be non-confusable otherwise.
A set of strings $ \C \subseteq \A_q^* $ is said to be a \emph{zero-error}
code \cite{shannon} for a given channel if every two different codewords
$ \myx, \myy \in \C $ are non-confusable.
In other words, the requirement is that the code $ \C $ is capable of correcting
\emph{all} possible patterns of errors in a given channel.

For technical reasons (see Remark \ref{rem:arm} ahead), we shall consider
codes whose codewords are of length $ \leq\! n $, rather than being exactly
equal to $ n $, and whose first symbol is non-zero.
More precisely, codes will be defined in the space
\begin{equation}
\label{eq:S}
  \Sp_q(n)  \defeq  \big( \A_q \!\setminus\! \{0\} \big) \times \bigcup_{i=0}^{n-1} \A_q^i .
\end{equation}
For any such code $ \C $, its rate is defined in the usual way as
$ \frac{1}{n} \log |\C| $, where $ \log $ is the logarithm to the base $ 2 $.
The zero-error capacity of a channel is the $ \limsup_{n\to\infty} $ of the
rates of optimal zero-error codes in $ \Sp_q(n) $.
The zero-error capacity of the $ q $-ary $ (\ell, r) $-duplication channel
is denoted by $ C_0^{\textnormal{dupl}}(q, \ell, r) $.
Note that the convention of defining codes in $ \Sp_q(n) $, rather than in,
e.g., $ \A_q^n $, does not affect the asymptotically achievable rates, i.e.,
the capacity.
It is adopted solely for convenience as it makes the discussion and the proofs
``cleaner''.

Another point that should be mentioned in this context is the following.
While zero-error codes are defined here by the (usual) requirement that
every two different codewords are non-confusable, in the setting where
multiple codewords are to be transmitted in succession over a channel
another issue arises that must be resolved in order for a code to be called
``zero-error''.
Namely, due to the facts that the codewords have possibly different lengths,
and that the duplication channel changes the lengths of the transmitted sequences,
it is not a priory clear how the receiver will be able to delimit the output
sequences that correspond to different codewords, and to decode them correctly.
In Section~\ref{sec:main} this problem is ignored and it is assumed that
only one codeword is transmitted per session.
This assumption is justified in some (DNA-based) data storage scenarios
where the stored codewords are physically separated.
The issue with the consecutive transmission of multiple codewords is discussed
briefly in Section \ref{sec:cw}, and a simple solution is presented.

\subsection{Related Work and Main Results}
 
The binary $ 0 $-insertion channel was first studied in \cite{levenshtein},
where a construction of codes correcting $ t $ insertions of zeros was presented
and asymptotic bounds on the cardinality of optimal codes derived, for any fixed
$ t $ and $ n \to \infty $.
No restrictions on the locations of inserted zeros were imposed in \cite{levenshtein},
so one might refer to this channel as the $ (1, \infty) $-$ 0 $-insertion channel
(i.e., $ r = \infty $).
Another construction for the same model was later given in \cite{dolecek+anantharam}.
Generalizations of the constructions from \cite{levenshtein} and \cite{dolecek+anantharam}
to the case $ \ell > 1 $ were given in \cite{kovacevic+tan2} and \cite{lenz2},
respectively.
The best known asymptotic bounds on codes correcting $ t $ insertions of blocks
$ 0^\ell $, for fixed $ t $ and $ n \to \infty $, were reported in \cite{kovacevic+tan2},
where the upper bound from \cite{levenshtein} (for $ \ell = 1 $) was improved for
every $ t > 2 $, and the bounds were also generalized to the case $ \ell > 1 $.

A construction of codes correcting \emph{all} possible error patterns in the
$ (\ell, \infty) $-duplication channel, and the consequent characterization of
its \emph{zero-error} capacity, were given in \cite{jain2}.
We note, however, that for $ \ell = 1 $, which is the main motivating model in
the present paper, the zero-error capacity of the $ (1, \infty) $-duplication
channel is trivially $ \log(q-1) $, so the mentioned results from \cite{jain2}
are only interesting for $ \ell > 1 $.
The zero-error capacity of a model that can be seen as the \emph{continuous-time}
version of the $ (1, r) $-duplication channel was determined in \cite{yeung}.
Namely, in \cite{yeung} a continuous-time channel was studied in which the input
waveforms are $ \A_q $-valued step functions, and in which the duration of each
``symbol'' $ a \in \A_q $ (the length of the interval in which the input waveform
has a fixed value $ a \in \A_q $) is changed by a random multiplicative factor.
This is analogous to our discrete-time $ (1,r) $-duplication channel in which
the length of an input run of identical symbols may change from $ u $ to $ v $,
where $ u \leq v \leq u (r + 1) $.

Our main results are a construction of optimal zero-error codes and a
characterization of the zero-error capacity of the $ (\ell, r) $-duplication
channel, for any fixed $ \ell $ and $ r $.
The method of construction is very similar to those given in \cite{jain2} and
\cite{yeung}, and all of them can in fact be seen as straightforward extensions
of the methods Shannon developed for DMCs \cite{shannon}.
Thus the above-mentioned results from \cite{jain2} can be recovered as a special
case of our results by letting $ r \to \infty $, while in another special case
($ \ell = 1 $) our results represent discrete-time analogs of those from \cite{yeung}.
In addition to being optimal in terms of their cardinality, the constructed
codes have a very simple combinatorial structure based on which an efficient
decoding algorithm is obtained.
Moreover, and further generalizing our findings, we obtain a characterization
of the \emph{constant-weight} zero-error capacity of the $ (\ell, r) $-$ 0 $-insertion
channel.

Finally, we mention in this context several more works on duplication channels
that are information theoretic in nature, but are concerned with different
problems.
The Shannon capacity of duplication channels---the largest rate achievable
in the vanishing-error, rather than the zero-error regime---was studied
extensively \cite{cheraghchi+ribeiro, drinea, iyengar, kirsch, mercier, mitzenmacher, ramezani}.
Tight bounds on the capacity were obtained in these works, but the exact
value remains elusive even for the most basic models.
The expressive power of duplications as a generative process was studied
in the recent works \cite{elishco, farnoud, jain}.

\section{Error-Free Communication Over\\Duplication Channels}
\label{sec:main}

This section presents a solution to the problem of error-free communication
over the $ (\ell, r) $-$ 0 $-insertion channel.
More precisely, a construction of optimal zero-error codes, a description of
the corresponding decoding algorithm, and a characterization of the zero-error
capacity are given.
Properties and numerical values of the capacity as a function of channel
parameters are also examined.

\subsection{Optimal Zero-Error Codes}
\label{sec:codes}

Let $ \B_{q, \ell, r} $ be the set of all finite $ q $-ary strings consisting
of a non-zero symbol followed by a block of zeros of length
\begin{equation}
\label{eq:runs}
\begin{aligned}
  &\frac{(r i + 1)(r \ell + 1)^j - 1}{r} - 1  =  \left(i+\frac{1}{r}\right)(r \ell + 1)^j - \frac{1}{r} - 1   \\
	&= i(r\ell+1)^j + \ell \left( (r\ell+1)^{j-1} + (r\ell+1)^{j-1} + \cdots + 1 \right) - 1 ,
\end{aligned}
\end{equation}
for some $ i \in \{1, 2, \ldots, \ell\} $ and $ j \geq 0 $, that is
\begin{equation}
\begin{aligned}
\label{eq:B}
  &\B_{q, \ell, r}   \\
	&\defeq \left\{ \sigma\,0^{\frac{(r i + 1)(r \ell + 1)^j - 1}{r} - 1} : 1 \leq \sigma \leq q-1, 1 \leq i \leq \ell, j \geq 0 \right\} .
\end{aligned}
\end{equation}
In the important special case $ \ell = 1 $ this reduces to
\begin{align}
\label{eq:B1}
  \B_{q, 1, r} 	=  \left\{ \sigma\,0^{\frac{(r + 1)^j - 1}{r} - 1} : 1 \leq \sigma \leq q-1, j \geq 1 \right\} ,
\end{align}
and further specializing to $ q = 2 $, $ r = 1 $ we get
\begin{equation}
\label{eq:B2}
  \B_{2, 1, 1}  =  \left\{ 1\,0^{2^j-2} : j \geq 1 \right\} .
\end{equation}
Note that the lengths of the blocks from $ \B_{q, \ell, r} $ form a geometric
progression (shifted for $ 1/r $) for every fixed $ i $, and that these lengths
are congruent to $ i $ modulo $ \ell $ (see \eqref{eq:runs}).

Define the code $ \C_{q, \ell, r}(n) $ as the set of all strings of length
$ \leq\!n $ that are composed of blocks from $ \B_{q, \ell, r} $, namely
\begin{equation}
\label{eq:C}
  \C_{q, \ell, r}(n)  \defeq  \B_{q, \ell, r}^* \cap \Sp_q(n) ,
\end{equation}
where $ \Sp_q(n) $ was defined in \eqref{eq:S}.
Theorem \ref{thm:codes} ahead claims that this code is an optimal zero-error
code for the $ q $-ary $ (\ell, r) $-$ 0 $-insertion channel.
Before we state and prove it, we describe first the intuition behind the
construction and explain how the set of constituent blocks from \eqref{eq:B}
arises as a solution (see also the discussion on constant-weight codes in
Section~\ref{sec:cw}, Figure~\ref{fig:code} in particular).

Recall that, for a transmitted sequence $ \myx = \sigma_1\,0^{u_1} \,\cdots\, \sigma_w\,0^{u_w} $,
$ \sigma_i \neq 0 $, the $ 0 $-insertion channel acts on the component blocks
$ \sigma_i 0^{u_i} $ independently of each other (see Section \ref{sec:channel}).
One may therefore attempt to find an optimal zero-error code $ \B_{q, \ell, r} $
in the set of all possible blocks $ \{ \sigma\,0^u : \sigma \in \A_q\!\setminus\!\{0\}, u \geq 0 \} $,
and then construct a code $ \C_{q, \ell, r} $ for the $ (\ell, r) $-$ 0 $-insertion
channel by concatenating the blocks from $ \B_{q, \ell, r} $ (i.e., define the
code in the set $ \B_{q, \ell, r}^* $).
As it turns out, this approach is optimal.

\begin{figure*}
\centering
  \includegraphics[width=0.85\textwidth]{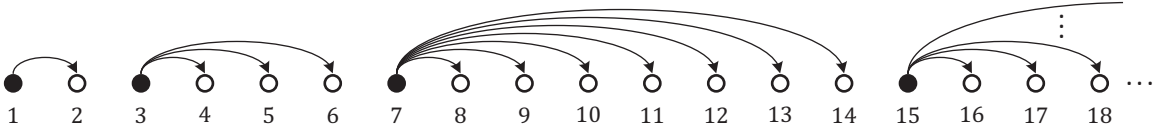}
\caption{Construction of the code $ \B_{2, 1, 1} \subseteq \{ 1\,0^u : u\geq 0\}$.
         The blocks $ 1\,0^{u} $ are represented by their lengths $ u+1 \in \{1, 2, \ldots \} $;
         codewords are depicted as black dots; and sequences that a particular codeword
				 can produce at the output of the $ (1, 1) $-$ 0 $-insertion channel are indicated
				 with the corresponding arrows.}%
\label{fig:constr}
\end{figure*}%

To find an optimal zero-error code $ \B_{q, \ell, r} $, we employ a simple
greedy algorithm.
First note that the channel preserves both the initial non-zero symbol and the
length of the input sequence modulo $ \ell $ (as it only adds blocks $ 0^\ell $),
so we can independently construct codes in each subset of blocks
$ \{ \sigma\,0^u : u+1 \equiv i  \pmod \ell \} $, defined for any fixed
$ \sigma \in \{1, \ldots, q-1\} $ and $ i \in \{1, \ldots, \ell\} $.
Now, for each of these $ (q-1)\ell $ subsets we do the following.
List the elements of a subset according to their length.
Select the first element on the list as a codeword, and delete from the list
all sequences that this codeword can produce at the output of the
$ (\ell, r) $-$ 0 $-insertion channel.
Then select the first element on the remaining list as a codeword, delete all
sequences that it can produce at the output of the channel, and so on.
Figure~\ref{fig:constr} illustrates this procedure for $ q = 2 $, $ \ell = 1 $,
$ r = 1 $ (there is only $ (q-1)\ell = 1 $ list in this case).
The set of sequences that $ 1\,0^u $ can produce at the output of the
$ (1, 1) $-$ 0 $-insertion channel is $ \{ 1\,0^v : u \leq v \leq 2u+1 \} $,
and so the resulting set of codewords is
$ \B_{2, 1, 1} = \big\{ 1, 1\,0^2, 1\,0^6, 1\,0^{14}, \ldots \big\} = \big\{ 1\,0^{2^j - 2} : j \geq 1 \big\} $.
For general parameters $ q, \ell, r $, the set of sequences that $ \sigma\,0^u $
can produce at the channel output is
$ \big\{ \sigma\,0^{u}, \sigma\,0^{u + \ell}, \ldots, \sigma\,0^{u + (u + 1)r\ell} \big\} $
(up to $ r $ blocks $ 0^\ell $ are inserted after each of the $ (u + 1) $ symbols
of the input sequence).
By using this fact it is not difficult to see that, in the subset of blocks
$ \{ \sigma\,0^u : u+1 \equiv i  \pmod \ell \} $, the set of codewords obtained
by the procedure just described is
$ \big\{ \sigma\,0^{i-1}, \sigma\,0^{i(r\ell+1) + \ell-1}, \sigma\,0^{i(r\ell+1)^2 + \ell(r\ell+1) + \ell-1}, \ldots \big\}
  = \big\{ \sigma\,0^{i(r\ell+1)^j + \ell  \left((r\ell+1)^{j-1} + (r\ell+1)^{j-1} + \cdots + 1 \right) -1} : j \geq 0 \big\} $.
The union of these sets over all $ i $ and $ \sigma $ is precisely the code
$ \B_{q, \ell, r} $ defined in \eqref{eq:B}.

\begin{theorem}
\label{thm:codes}
$ \C_{q, r, \ell}(n) $ is a zero-error code for the $ (\ell, r) $-$ 0 $-insertion
channel.
Moreover, every zero-error code $ \C \subseteq \Sp_q(n) $ for the
$ (\ell, r) $-$ 0 $-insertion channel satisfies $ |\C| \leq |\C_{q, r, \ell}(n)| $.
\end{theorem}
\begin{IEEEproof}
To prove the first part of the statement, we need to demonstrate that any two
distinct codewords from $ \C_{q, \ell, r}(n) $ are non-confusable in the
$ (\ell, r) $-$ 0 $-insertion channel.
Since all codewords are composed of blocks from $ \B_{q, \ell, r} $, and since
the channel acts independently on the component blocks of a codeword, it is
enough to show that any two distinct blocks from $ \B_{q, \ell, r} $ are
non-confusable.
So consider two such blocks,
$ \myx = \sigma\,0^{\frac{(r i + 1)(r \ell + 1)^j - 1}{r} - 1} $ and
$ \myy = \sigma\,0^{\frac{(r k + 1)(r \ell + 1)^m - 1}{r} - 1} $.
There are two cases to examine:
\begin{inparaenum}
\item[1.)]
$ i \neq k $, and
\item[2.)]
$ i = k $ but $ j < m $ (the case $ j > m $ will follow by symmetry).
\end{inparaenum}
In the first case, since $ |\myx| \equiv i  \pmod \ell $ and $ |\myy| \equiv k  \pmod \ell $,
we have $ |\myx| \not\equiv |\myy| \pmod \ell $.
Since the channel preserves the lengths of strings modulo $ \ell $, we
conclude that $ \myx $ and $ \myy $ cannot produce strings of the same
length at the channel output and are therefore non-confusable.
In the second case, since the maximum number of inserted zeros in $ \myx $
is $ |\myx| r \ell $, the maximum length of the resulting output string is
$
  |\myx| + |\myx| r \ell  = \linebreak
	\frac{ (r i + 1)(r \ell + 1)^j - 1 }{ r } (r \ell + 1)  =
  \frac{ (r i + 1)(r \ell + 1)^{j+1} - 1 }{ r } - \ell  <
  \frac{ (r i + 1)(r \ell + 1)^{m} - 1 }{ r }  =
  |\myy| .
$
As the channel can only increase the length of the transmitted string, we
conclude that $ \myy $ can never produce the same strings at the channel output
that $ \myx $ can, and therefore, $ \myx $ and $ \myy $ are non-confusable.
This proves that $ \B_{q, \ell, r} $ is a zero-error code for the $ (\ell, r) $-0-insertion
channel, which further implies that $ \C_{q, \ell, r}(n) $ is a zero-error code
as well.

We now prove the second part of the statement which claims that the code
$ \C_{q, \ell, r}(n) $ is optimal.
Let $ f : \Sp_q(n) \to \Sp_q(n) $ be a mapping defined as follows:
for any string $ \myx \in \Sp_q(n) $, $ f(\myx) $ is obtained by removing from
each run of zeros in $ \myx $ as many blocks $ 0^\ell $ as is necessary and
sufficient in order to obtain a run of zeros whose length is of the form
$ \frac{ (r i + 1)(r \ell + 1)^j - 1 }{ r } - 1 $, with
$ 1 \leq i \leq \ell $ and $ j \geq 0 $.
We claim that $ f $ has the following property: if two strings $ \myx, \myy $
are non-confusable in the $ (\ell, r) $-0-insertion channel, then their
images $ f(\myx), f(\myy) $ are non-confusable as well.
Again, in order to show this, we may assume without loss of generality that
$ \myx $ and $ \myy $ are of the form $ \myx = \sigma\,0^{u} $, $ \myy = \sigma\,0^{v} $,
for some $ \sigma \in \A_q \setminus \{0\} $, in which case we have
$ f(\myx) = \sigma\,0^{\frac{(r i + 1)(r \ell + 1)^j - 1}{r} - 1} $,
$ f(\myy) = \sigma\,0^{\frac{(r k + 1)(r \ell + 1)^m - 1}{r} - 1} $.
Let also $ |\myx| < |\myy| $.
Now, suppose that $ f(\myx), f(\myy) $ are \emph{confusable}.
Then, since $ f(\myx), f(\myy) \in \B_{q,\ell,r} $, it follows from the
first part of the proof that we must have $ f(\myx) = f(\myy) $, i.e,
$ (i, j) = (k, m) $.
By the definition of $ f $, we then conclude that $ |\myx| \equiv |\myy| \equiv i \pmod \ell $
(because $ f $ removes blocks $ 0^\ell $ from its argument and hence preserves
its length modulo $ \ell $), and that
$ \frac{(r i + 1)(r \ell + 1)^j - 1}{r}  \leq
	|\myx|  <
	|\myy|  \leq
	\frac{(r i + 1)(r \ell + 1)^{j+1} - 1}{r} - \ell
$
(where the last inequality holds because assuming otherwise, namely
$ |\myy| > \frac{(r i + 1)(r \ell + 1)^{j+1} - 1}{r} - \ell $, and using the
fact that $ |\myy| \equiv i \pmod \ell $, would imply that
$ |\myy| \geq \frac{(r i + 1)(r \ell + 1)^{j+1} - 1}{r} $, and hence $ m \geq j + 1 $,
contradicting the above-established fact that $ m = j $).
This further implies that $ |\myy| - |\myx| $ is a multiple of $ \ell $, and that
$ |\myy| - |\myx|  \leq  \linebreak
  (r i + 1)(r \ell + 1)^j \ell - \ell  =
	\frac{(r i + 1)(r \ell + 1)^j - 1}{r} r \ell  \leq
	|\myx| r \ell
$.
This means that the string $ \myy $ can be obtained from the string $ \myx $
by inserting $ |\myx| r $ blocks $ 0^\ell $ in the latter, and therefore
$ \myx $ and $ \myy $ are confusable in the $ (\ell, r) $-0-insertion channel.
We have thus established the claimed property of $ f $: if $ f(\myx) $ and
$ f(\myy) $ are confusable, then so are $ \myx $ and $ \myy $.
Now let $ \C \subseteq \Sp_q(n) $ be an arbitrary zero-error code for the
$ (\ell, r) $-0-insertion channel, and denote by $ f(\C) $ the image of
$ \C $ under the mapping $ f $.
We first notice that $ f(\C) $ is itself a zero-error code, because
$ f(\C) \subseteq \C_{q,\ell,r}(n) $ by the definition of $ f $.
Moreover, $ |f(\C)| = |\C| $, i.e., $ f $ is injective over any zero-error
code $ \C $.
This follows from the above-mentioned property of $ f $: since any two different
codewords $ \myx, \myy \in \C $ are non-confusable, their images $ f(\myx), f(\myy) $
are necessarily non-confusable as well and, in particular, $ f(\myx) \neq f(\myy) $.
To sum up, what we have just shown is that \emph{every} zero-error code
$ \C \subseteq \Sp_q(n) $ for the $ (\ell, r) $-0-insertion channel can be
bijectively mapped to a subcode of $ \C_{q,\ell,r}(n) $, implying that
$ |\C| \leq |\C_{q, r, \ell}(n)| $.
\end{IEEEproof}

\begin{remark}[Shannon's adjacency reducing mappings]
\label{rem:arm}
\textnormal{
The mappings having the property that the images of two non-confusable inputs
are themselves non-confusable, such as the one from the preceding proof, were
introduced by Shannon under the name ``adjacency reducing mappings'' \cite{shannon}.
It was shown in \cite[Thm~3]{shannon} that if, for a DMC with input alphabet
$ \A_q $, an adjacency reducing mapping $ f : \A_q \to \A_q $ exists such that
all letters in $ f(\A_q) $ are non-confusable, then the zero-error capacity
of that DMC equals $ \log |f(\A_q)| $.
This result does not directly apply in our case because the DMC to which the
$ (\ell, r) $-$ 0 $-insertion channel is equivalent has an infinite alphabet,
as well as costs assigned to the letters in this alphabet (see Section \ref{sec:channel}).
However, one can see from the proof of Theorem~\ref{thm:codes} that the main
idea is still the same and that only several modifications to the method from
\cite{shannon} are needed.
For example, in our setting the images $ f(\myx) $ do not all have the same
length (or ``cost''), even if the originals $ \myx $ do.
This is why we have assumed that codeword lengths are \emph{upper-bounded} by
$ n $, as opposed to being exactly equal to $ n $ (see \eqref{eq:S} and \eqref{eq:C}).
\myqed
}
\end{remark}

\begin{remark}[Optimal codes for the duplication channel]
\label{rem:assmp}
\textnormal{
Recall from Section~\ref{sec:channel} that the $ (\ell, r) $-duplication
channel is equivalent to a channel which inserts up to $ r $ blocks
$ 0^\ell $ after every symbol of the input sequence \emph{except} for the
first $ \ell-1 $ symbols.
Recall also that, in order to simplify the discussion, we have adopted
the following conventions in our definition of the $ (\ell, r) $-$ 0 $-insertion
channel:
\begin{inparaenum}
\item[(a)]
up to $ r $ blocks $ 0^\ell $ may be inserted after \emph{every} symbol of
the input sequence, and
\item[(b)]
the first symbol of the transmitted sequence is always non-zero.
\end{inparaenum}
It is clear that these assumptions have no effect on the asymptotic analysis
and were adopted for convenience.
In other words, dropping these assumptions would not change the zero-error
capacity of the channel.
However, if one wishes to study strict optimality of codes in the finite-blocklength
regime, then they must be taken into account.
}

\textnormal{
We wish to point out here that optimal codes can easily be constructed by
similar methods even if the above assumptions are dropped.
Namely, an optimal code of length $ \leq n $ for the channel which inserts
up to $ r $ blocks $ 0^\ell $ after every symbol of the input sequence except
for the first $ \ell-1 $ symbols (and which has no restriction on the first
symbol of a codeword) is given by
\begin{align}
  \C'_{q, r, \ell}(n) = \left( \A_q^\ell \times \Bb_{q, r, \ell} \times \B^*_{q, r, \ell} \right) \cap \bigcup_{i=0}^{n} \A_q^i ,
\end{align}
where
\begin{align}
  \Bb_{q, \ell, r} \defeq  \left\{ 0^{\frac{(r i + 1)(r \ell + 1)^j - 1}{r} - 1} : 1 \leq i \leq \ell, j \geq 0 \right\} .
\end{align}
(Note that $ \Bb_{q, \ell, r} $ is $ \B_{q, \ell, r} $ with the first symbol
of every block omitted, i.e., $ \B_{q, \ell, r} = (\A_q \setminus \{0\}) \times \Bb_{q, \ell, r} $.)
In words, the prefix of length $ \ell $ and all the non-zero symbols after
that prefix are chosen arbitrarily in every codeword (as they are left intact
by the channel), and the lengths of all runs of zeros after the prefix are
required to be of the form \eqref{eq:runs}.
\myqed
}
\end{remark}

\subsection{Decoding Algorithm}
\label{sec:decoder}

Proof of Theorem \ref{thm:codes} also outlined a simple linear-time decoding
algorithm for the codes $ \C_{q, \ell, r}(n) $, which we state here explicitly.
To recover the transmitted string from the received string $ \bs{z} $, the
decoder first divides $ \bf{z} $ into blocks of the form $ \sigma\,0^{u} $,
$ \sigma \neq 0 $, and then applies the function $ f $ to each of the blocks.
In other words, the decoder shortens each run of zeros in $ \bs{z} $ by
removing as many blocks $ 0^\ell $ as needed so that the length of the resulting
run is the largest integer of the form
$ \frac{(r i + 1)(r \ell + 1)^j - 1}{r} - 1 $, $ 1 \leq i \leq \ell $, $ j \geq 0 $.
In mathematically precise terms, the decoder's output is
\begin{equation}
  f(\sigma\,0^u) = \sigma\,0^{\frac{(r i + 1)(r \ell + 1)^j - 1}{r} - 1} ,
\end{equation}
where $ i $ is the unique integer in $ \{1, \ldots, \ell\} $ satisfying
\begin{equation}
  i \equiv u+1 \pmod \ell ,
\end{equation}
and $ j $ is computed as
\begin{equation}
  j = \left\lfloor \log_{r\ell+1}\!\left( \frac{r(u+1) + 1}{ri+1} \right) \right\rfloor .
\end{equation}

\subsection{Zero-Error Capacity}
\label{sec:cap}

Define the function
\begin{equation}
\label{eq:v}
  v_{q, \ell, r}(x)  \defeq  (q-1) \sum_{j=0}^{\infty} \sum_{i=1}^{\ell}  x^{ \frac{ (r i + 1)(r \ell + 1)^j - 1}{r} } .
\end{equation}
In the special case $ \ell = 1 $ this reduces to
\begin{equation}
\label{eq:v1}
  v_{q, 1, r}(x)  =  (q-1)  \sum_{j=1}^{\infty}  x^{ \frac{ (r + 1)^j - 1}{r} } ,
\end{equation}
and further specializing to $ q = 2 $, $ r = 1 $ we get%
\footnote{The function $ v_{2, 1, 1}(x) $ (in fact, the related function
$ x (1 + v_{2, 1, 1}(x)) = \sum_{j=0}^{\infty} x^{2^j} $) is of interest
in transcendental number theory; see, e.g., \cite{adamczewski}.}
\begin{equation}
\label{eq:v2}
  v_{2, 1, 1}(x)  =  \sum_{j=1}^{\infty} x^{2^j - 1} .
\end{equation}

\begin{theorem}
\label{thm:cap}
The zero-error capacity of the $ (\ell, r) $-duplication channel is equal to
$ C_0^{\textnormal{dupl}}(q, \ell, r) = -\log\rho $, where $ \rho $ is the
unique positive solution to the equation $ v_{q, \ell, r}(x) = 1 $.
\end{theorem}
\begin{IEEEproof}
By Theorem \ref{thm:codes} and the equivalence of the duplication and the
$ 0 $-insertion channel (see Section \ref{sec:channel}), we conclude that the
zero-error capacity of the $ (\ell, r) $-duplication channel is equal to the
asymptotic rate of the codes $ \C_{q, \ell, r}(n) $, that is
$ C_0^{\textnormal{dupl}}(q, \ell, r) = \lim_{n \to \infty} \frac{1}{n} \log |\C_{q, \ell, r}(n)| $.

The following recurrence relation holds:
\begin{equation}
\begin{aligned}
\label{eq:Crec}
  &\big| \C_{q, \ell, r}(n) \big|   \\
  &=  (q-1) \sum_{j=0}^{\infty} \sum_{i=1}^{\ell}  \left|\C_{q, \ell, r}\!\left( n - \frac{(r i + 1)(r \ell + 1)^j - 1}{r} \right)\right| ,
\end{aligned}
\end{equation}
which is seen by considering which of the blocks from $ \B_{q, \ell, r} $ is
the first block in each particular codeword of $ \C_{q, \ell, r}(n) $.
The theorem is then obtained by a standard application of the results of analytic
combinatorics \cite{flajolet} after noting that the characteristic equation of
the relation \eqref{eq:Crec} is
\begin{equation}
\label{eq:char}
  1 - v_{q, \ell, r}(x) = 0 .
\end{equation}
Namely, the exponential growth rate of a quantity defined by a linear recurrence
with characteristic equation of the form \eqref{eq:char} is determined by the
complex root of that equation that is closest to the origin.
It is known \cite[Sec.~IV.3]{flajolet} that equations of the form \eqref{eq:char}
(where the function $ v_{q, \ell, r}(x) $ has only positive coefficients) have a unique
positive root and that this root is necessarily closest to the origin.
\end{IEEEproof}
\vspace{2mm}

We note that the statements \cite[Thm~16 and Cor.~18]{jain2}, which characterize
the optimal zero-error codes and the zero-error capacity for the
$ (\ell, \infty) $-duplication channel, can be recovered as a special case
of Theorems \ref{thm:codes} and \ref{thm:cap} above.
To see this just observe that, as $ r \to \infty $, \eqref{eq:B} transforms to
\begin{equation}
\label{eq:Binfty}
  \B_{q, \ell, \infty}  \defeq  \Big\{ \sigma\,0^{i} : 1 \leq \sigma \leq q-1, 0 \leq i \leq \ell-1 \Big\} ,
\end{equation}
and \eqref{eq:v} to
\begin{equation}
\label{eq:vinfty}
  v_{q, \ell, \infty}(x)  \defeq  (q-1)  \sum_{i=1}^{\ell} x^{i} .
\end{equation}
Thus, the set of all sequences that have no run of zeros of length $ \geq \ell $
($ (0, \ell-1) $-constrained sequences)
is an optimal
zero-error code for the $ (\ell, r) $-$ 0 $-insertion channel \cite{jain2}.

We also note that, in another special case ($ \ell = 1 $; see \eqref{eq:B1} and \eqref{eq:v1}),
Theorems \ref{thm:codes} and \ref{thm:cap} represent discrete-time analogs
of the results reported in \cite{yeung}.

\subsection{Properties and Numerical Values of the Capacity}

The following proposition shows how the zero-error capacity $ C_0^{\textnormal{dupl}}(q, \ell, r) $
behaves as a function of its parameters.
The statement is intuitively clear so we omit the formal proof
(for example, increasing $ r $ means that the channel is getting ``noisier'',
so the resulting capacity must be smaller).

\begin{proposition}
\label{thm:cap_prop}
The function $ C_0^{\textnormal{dupl}}(q, \ell, r) $ has the following properties:
\begin{itemize}
\item[(a)]
For any fixed $ \ell, r $, it is monotonically increasing in $ q $, with
$ \lim_{q \to \infty}  \big( C_0^{\textnormal{dupl}}(q, \ell, r) - \log q \big) = 0 $.
\item[(b)]
For any fixed $ q, r $, it is monotonically increasing in $ \ell $, with
$ \lim_{\ell \to \infty}  C_0^{\textnormal{dupl}}(q, \ell, r)  =  \log q $.
\item[(c)]
For any fixed $ q, \ell $, it is monotonically decreasing in $ r $, with
$ C_0^{\textnormal{dupl}}(q, \ell, \infty)  \defeq  \lim_{r \to \infty}  C_0^{\textnormal{dupl}}(q, \ell, r)  = -\log \rho_\infty $.
Here $ \rho_\infty $ is the unique positive solution to $ v_{q, \ell, \infty}(x) = 1 $,
and $ v_{q, \ell, \infty}(x) $ is defined in \eqref{eq:vinfty}.
In particular, $ C_0^{\textnormal{dupl}}(q, 1, \infty) = \log(q-1) $.
\end{itemize}
\end{proposition}

While in most cases the root $ \rho $ and the capacity $ -\log\rho $ are not
(and most likely cannot be) given in an explicit form, their values can easily
be computed/approximated numerically.
Since the \emph{powers} in the sum \eqref{eq:v} grow exponentially fast in $ j $,
if one wishes to compute the first $ d $ digits of $ \rho $, then it is
sufficient to approximate the infinite sum in \eqref{eq:char} with its first
$ \sim \log d $ summands and find the positive root of the resulting polynomial.
Moreover, this root is guaranteed to be in the range $ \big[\frac{1}{q}, \frac{1}{q-1}\big] $,
which further simplifies the numerical procedures for finding it.
For example, for the binary $ (1, 1) $-duplication channel we get
$ \rho \approx 0.659 $ and $ C_0^{\textnormal{dupl}}(2, 1, 1) = -\log \rho \approx 0.602 $
bits per symbol.

\begin{figure}
 \centering
  \subfigure[$ q = 2 $, $ \ell = 1, 2, 3, 4 $ (the lowest curve corresponds to $ \ell = 1 $).]
  {
   \includegraphics[width=\columnwidth]{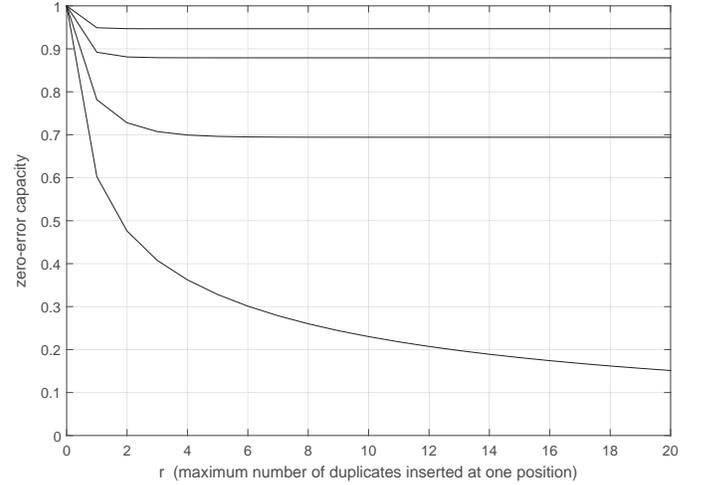}
   \label{fig:plotq2}
  }
	\vspace{1mm}
  \subfigure[$ q = 4 $, $ \ell = 1, 2, 3, 4 $ (the lowest curve corresponds to $ \ell = 1 $).]
  {
	 \makebox[\columnwidth]{
    \includegraphics[width=\columnwidth]{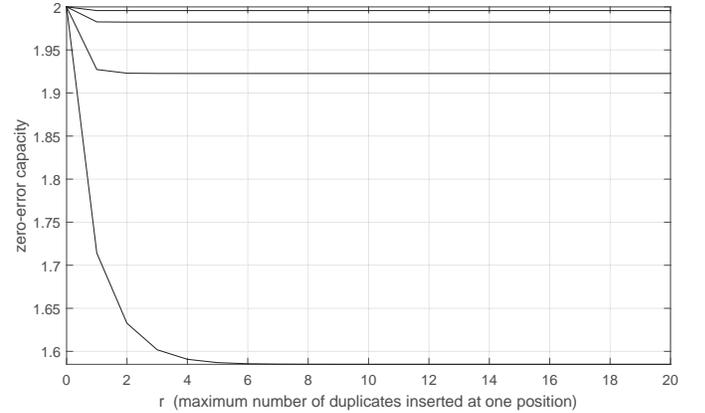}
    \label{fig:plotq4}
	 }
	}
\caption{Zero-error capacity of the $ q $-ary $ (\ell,r) $-duplication channel as a
         function of the parameter $ r $. The plots are given for the binary and quaternary
         alphabets, and for four different values of the duplication length $ \ell $.}
\label{fig:plot}
\end{figure}%

As an illustration, the zero-error capacity $ C_0^{\textnormal{dupl}}(q, \ell, r) $
is plotted in Figure~\ref{fig:plot} as a function of $ r $, for several values
of $ q $ and $ \ell $.
We make here a few observations about this function based on the obtained numerical
results.
For most values of the parameters $ q $ and $ \ell $, the value of
$ C_0^{\textnormal{dupl}}(q, \ell, r) $ very quickly converges to
$ C_0^{\textnormal{dupl}}(q, \ell, \infty) $ as $ r $ grows.
The convergence is slowest for the binary sticky-insertion channel (for which the
limiting value is $ C_0^{\textnormal{dupl}}(2, 1, \infty) = 0 $), and it becomes
faster as either $ q $ or $ \ell $ (or both) increase.
For example, for \emph{every} $ (q, \ell) $ with $ q \geq 3 $ and $ \ell \geq 3 $,
the value of $ C_0^{\textnormal{dupl}}(q, \ell, r) $ practically drops to its
asymptotic value $ C_0^{\textnormal{dupl}}(q, \ell, \infty) $ already at $ r = 1 $.
The exact value of the difference
$ C_0^{\textnormal{dupl}}(q, \ell, r) - C_0^{\textnormal{dupl}}(q, \ell, \infty) $,
which represents the penalty incurred by assuming that an unbounded number of
duplicates may be inserted at a single position (as in the model in \cite{jain2})
when in fact this number is bounded by $ r $, can by Theorem~\ref{thm:cap} and
Proposition~\ref{thm:cap_prop} be computed as
$ \log(\rho_\infty/\rho) $, where $ \rho $ (resp. $ \rho_\infty $) is the positive
solution to $ v_{q,\ell,r}(x) = 1 $ (resp. $ v_{q,\ell,\infty}(x) = 1 $).
Also note that, for fixed $ q $ and $ r $, the value of the capacity quickly
converges to its trivial upper bound $ \log q $ as $ \ell $ grows, meaning that
long duplication errors (large $ \ell $) can be corrected by using codes with
negligible redundancy.

\section{Constant-Weight Codes for the\\$ 0 $-Insertion Channel}
\label{sec:cw}

\begin{figure}
\centering
  \includegraphics[width=\columnwidth]{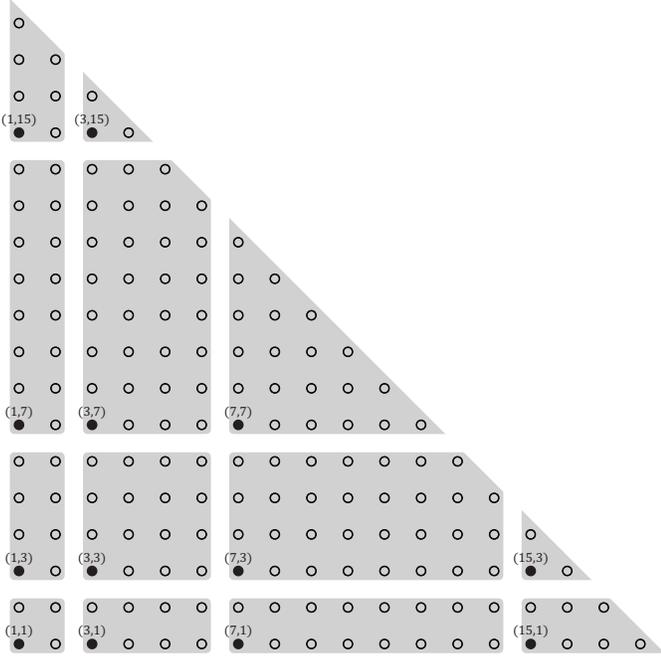}
\caption{The space $ \Sp_2(19; 2) $ (the set of all binary sequences of length $ \leq 19 $
         and weight $ 2 $ that begin with a $ 1 $) and the code $ \C_{2, 1, 1}(19; 2) $.
         Codewords are depicted as black dots, and a point $ (u, v) $ represents
				 the sequence $ 1 0^{u-1} 1 0^{v-1} $
				 (thus the two coordinates $ u $ and $ v $ are the lengths of the constituent blocks).
				 The set of sequences that a particular codeword can produce at the output of the
				 $ (1, 1) $-0-insertion channel is represented by the corresponding gray region.
         Note that $ | \C_{2, 1, 1}(19; 2) | = 13 $.
				 In contrast, the corresponding code for the channel with $ r = \infty $
				 has only one codeword (the binary sequence $ 11 $), $ | \C_{2, 1, \infty}(19; 2) | = 1 $.}
\label{fig:code}
\end{figure}

In this section we extend the results presented above to the case where all
codewords are required to have the same Hamming weight.
The need to analyze constant-weight codes for the $ 0 $-insertion channel
arises naturally as this channel preserves the Hamming weight of the
transmitted sequence and, therefore, an optimal code is a union of optimal
constant-weight codes over all possible weights.
In particular, the subcode of $ \C_{q, \ell, r}(n) $ consisting of all codewords
of weight $ w $, denoted $ \C_{q, \ell, r}(n; w) $, is an optimal zero-error code
of length $ n $ and weight $ w $ for the $ (\ell, r) $-$ 0 $-insertion channel.
This code is illustrated in Figure \ref{fig:code} for $ q = 2 $, $ \ell = 1 $,
$ r = 1 $, $ n = 19 $, $ w = 2 $.

Constant-weight codes are interesting in the present context for another
reason -- they solve quite easily the issue with codeword concatenation, i.e.,
consecutive transmission of multiple codewords.
Namely, since the $ 0 $-insertion channel affects the length of the transmitted
sequence, if multiple codewords are being sent in succession the receiver may
not be able to infer the boundaries between the output sequences that correspond
to different codewords and to decode them correctly.
If one uses constant-weight codes of weight $ w $, however, the receiver can
recognize the boundaries by simply counting the non-zero symbols and partitioning
the output sequence into segments of the form
$ \sigma_1\,0^{u_1} \, \cdots \, \sigma_w\,0^{u_w} $, where $ \sigma_i \neq 0 $.

\begin{remark}
\textnormal{
We should emphasize that we are analyzing here constant-weight codes for the
$ 0 $-insertion channel, and \emph{not} constant-weight codes for the duplication
channel.
Recalling the transformation $ \phi_\ell $ that translates codes for
one into codes for the other (see Section \ref{sec:channel}), we see that a
constant-weight requirement in the $ (1,r) $-$ 0 $-insertion channel corresponds
to a constant-number-of-runs-of-identical-symbols requirement in the
$ (1,r) $-duplication channel.
\myqed
}
\end{remark}

For $ \omega \in [0, 1] $, let $ C_0^{\textnormal{$0$-ins}}(\omega; q, \ell, r) $
denote the largest rate achievable by zero-error codes of length $ n \to \infty $
and weight $ w \sim \omega n $ in the $ (\ell, r) $-$ 0 $-insertion channel,
i.e., the constant-weight zero-error capacity of the $ (\ell, r) $-$ 0 $-insertion
channel.
Let also $ C_0^{\textnormal{$0$-ins}}(q, \ell, r) $ be the zero-error capacity
of the $ (\ell, r) $-$ 0 $-insertion channel (with no restriction on the weight).

\begin{theorem}
\label{thmn:cwcap}
For $ \omega \in (0, 1) $, the constant-weight zero-error capacity of the
$ (\ell, r) $-$ 0 $-insertion channel is equal to
\begin{equation}
  C_0^{\textnormal{$ 0 $-ins}}(\omega; q, \ell, r)
	= \omega \log\!\left(\! (q-1)  \sum_{j=0}^{\infty} \sum_{i=1}^{\ell} \rho_{\omega}^{ \frac{ (r i + 1)(r \ell + 1)^j - 1}{r} - \frac{1}{\omega} } \!\right) ,
\end{equation}
where $ \rho_w $ is the unique positive solution to the equation
$ \sum_{j=0}^{\infty} \sum_{i=1}^{\ell} \Big( \frac{ (r i + 1)(r \ell + 1)^j - 1}{r} - \frac{1}{\omega} \Big) x^{ \frac{ (r i + 1)(r \ell + 1)^j - 1}{r} } = 0 $.
Also,
$ C_0^{\textnormal{$0$-ins}}(0; q, \ell, r) = 0 $ and
$ C_0^{\textnormal{$0$-ins}}(1; q, \ell, r) = \log(q-1) $.

For any fixed $ q, \ell, r $, $ C_0^{\textnormal{$ 0 $-ins}}(\omega; q, \ell, r) $
is a continuous, strictly concave function of $ \omega $.
It attains its maximal value at
\begin{align}
\nonumber
  &\omega^*   \\
	&= \left(\! (q-1) \sum_{j=0}^{\infty} \sum_{i=1}^{\ell}   \frac{ (r i + 1)(r \ell + 1)^j - 1}{r} \rho^{ \frac{ (r i + 1)(r \ell + 1)^j - 1}{r} } \!\right)^{\!-1}
\end{align}
where $ \rho $ is the unique positive solution to $ v_{q, \ell, r}(x) = 1 $,
and\linebreak this value is
$ {C_0^{\textnormal{$ 0 $-ins}}(\omega^*; q, \ell, r) =
  C_0^{\textnormal{$ 0 $-ins}}(q, \ell, r) =
  C_0^{\textnormal{dupl}}(q, \ell, r)} $.%
\end{theorem}
\begin{IEEEproof}
By the observation from the first paragraph of this section we conclude that
$ C_0^{\textnormal{$ 0 $-ins}}(\omega; q, \ell, r) = \lim_{n \to \infty} \frac{1}{n} \log\!\big|\C_{q, \ell, r}(n; \omega n)\big| $.
Determining this limit, which represents the exponential growth rate of
$ \big|\C_{q, \ell, r}(n; \omega n)\big| $, is analogous to the derivation
of \cite[Lem.~1]{kovacevic}, which in turn is an application of the methods
described in \cite[Sec.~12.2]{pemantle+wilson}, so we omit the details.
The main difference with respect to \cite{kovacevic} is that the building
blocks of codewords are in the present case from $ \B_{q, \ell, r} $, while
in \cite{kovacevic} they are from $ \{ 1 \, 0^d, \ldots, 1 \, 0^k\} $.
\end{IEEEproof}
\vspace{2mm}

Therefore, the constant-weight codes $ \C_{q, \ell, r}(n; \omega^* n) $
achieve the zero-error capacity of the $ (\ell, r) $-$ 0 $-insertion channel.
For example, for $ q = 2 $, $ \ell = 1 $, $ r = 1 $, we have $ \rho \approx 0.659 $,
and the optimizing weight---the relative weight of zero-error-capacity-achieving
constant-weight codes---is $ \omega^* \approx 0.519 $.

\pagebreak
\section*{Acknowledgment}

The author would like to thank the referees, for their thorough reviews and
helpful comments on the original version of the manuscript, and Jo\~{a}o Ribeiro
(Imperial College London), for bringing \cite{cheraghchi+ribeiro} to his attention.

\IEEEtriggeratref{12}



\begin{thebibliography}{99}

\bibitem{adamczewski}
   B. Adamczewski,
   ``The Many Faces of the Kempner Number,''
   \emph{J. Integer Seq.}, vol.~16, article 13.2.15, 2013.
\bibitem{cheraghchi+ribeiro}
   M. Cheraghchi and J. Ribeiro,
   ``Sharp Analytical Capacity Upper Bounds for Sticky and Related Channels,''
   \emph{IEEE Trans. Inf. Theory}, to appear.
	 Published online at: \href{https://doi.org/10.1109/TIT.2019.2920375}{https://doi.org/10.1109/TIT.2019.2920375}.
\bibitem{dolecek+anantharam}
   L. Dolecek and V. Anantharam,
   ``Repetition Error Correcting Sets: Explicit Constructions and Prefixing Methods,''
   \emph{SIAM J. Discrete Math.}, vol.~23, no.~4, pp.~2120--2146, 2010.
\bibitem{drinea}
   E. Drinea and M. Mitzenmacher,
   ``Improved Lower Bounds for the Capacity of i.i.d. Deletion and Duplication Channels,''
   \emph{IEEE Trans. Inf. Theory}, vol.~53, no.~8, pp.~2693--2714, 2007.
\bibitem{elishco}
   O. Elishco, F. Farnoud, M. Schwartz, and J. Bruck,
   ``The Capacity of Some P\'{o}lya String Models,''
   preprint: \href{https://arxiv.org/abs/1808.06062v1}{arXiv:1808.06062v1}, Aug. 2018.
\bibitem{farnoud}
   F. Farnoud, M. Schwartz, and J. Bruck,
   ``The Capacity of String-Duplication Systems,''
   \emph{IEEE Trans. Inf. Theory}, vol.~62, no.~2, pp.~811--824, 2016.
\bibitem{flajolet}
   P. Flajolet and R. Sedgewick,
   \emph{Analytic Combinatorics},
   Cambridge University Press, 2009.
\bibitem{iyengar}
   A. R. Iyengar, P. H. Siegel, and J. K. Wolf,
   ``On the Capacity of Channels With Timing Synchronization Errors,''
   \emph{IEEE Trans. Inf. Theory}, vol.~62, no.~2, pp.~793--810, 2016.
\bibitem{jain}
   S. Jain, F. Farnoud, and J. Bruck,
   ``Capacity and Expressiveness of Genomic Tandem Duplication,''
   \emph{IEEE Trans. Inf. Theory}, vol.~63, no.~10, pp.~6129--6138, 2017.
\bibitem{jain2}
   S. Jain, F. Farnoud, M. Schwartz, and J. Bruck,
   ``Duplication-Correcting Codes for Data Storage in the DNA of Living Organisms,''
   \emph{IEEE Trans. Inf. Theory}, vol.~63, no.~8, pp.~4996--5010, 2017.
\bibitem{kirsch}
   A. Kirsch and E. Drinea,
   ``Directly Lower Bounding the Information Capacity for Channels With I.I.D. Deletions and Duplications,''
   \emph{IEEE Trans. Inf. Theory}, vol.~56, no.~1, pp.~86--102, 2010.
\bibitem{kovacevic}
   M. Kova\v{c}evi\'c,
   ``Runlength-Limited Sequences and Shift-Correcting Codes: Asymptotic Analysis,''
   \emph{IEEE Trans. Inf. Theory}, vol. 65, no. 8, pp. 4804--4814, 2019.
\bibitem{kovacevic+tan2}
   M. Kova\v{c}evi\'c and V. Y. F. Tan,
   ``Asymptotically Optimal Codes Correcting Fixed-Length Duplication Errors in DNA Storage Systems,''
   \emph{IEEE Commun. Lett.}, vol.~22, no.~11, pp.~2194--2197, 2018.
\bibitem{lenz2}
   A. Lenz, N. J\"{u}nger, and A. Wachter-Zeh,
   ``Bounds and Constructions for Multi-Symbol Duplication Error Correcting Codes,''
   preprint: \href{https://arxiv.org/abs/1807.02874v3}{arXiv:1807.02874v3}, Sep. 2018.
\bibitem{levenshtein}
   V. I. Levenshtein,
	 ``Binary Codes Correcting Deletions and Insertions of the Symbol $ 1 $'' (in Russian),
	 \emph{Probl. Peredachi Inf.}, vol.~1, no.~1, pp.~12--25, 1965.
\bibitem{levenshtein2}
   V. I. Levenshtein,
	 ``Binary Codes Capable of Correcting Deletions, Insertions, and Reversals,''
   \emph{Sov. Phys.--Dokl.}, vol. 10, no. 8, pp. 707--710, 1966.
\bibitem{mercier_survey}
   H. Mercier, V. K. Bhargava, and V. Tarokh,
   ``A Survey of Error-Correcting Codes for Channels With Symbol Synchronization Errors,''
   \emph{IEEE Commun. Surveys Tuts.}, vol.~12, no.~1, pp.~87--96, 2010.
\bibitem{mercier}
   H. Mercier, V. Tarokh, and F. Labeau,
	 ``Bounds on the Capacity of Discrete Memoryless Channels Corrupted by Synchronization and Substitution Errors,''
	 \emph{IEEE Trans. Inf. Theory}, vol.~58, no.~7, pp.~4306--4330, 2012.
\bibitem{mitzenmacher}
   M. Mitzenmacher,
	 ``Capacity Bounds for Sticky Channels,''
	 \emph{IEEE Trans. Inf. Theory}, vol.~54, no.~1, pp.~72--77, 2008.
\bibitem{mitzenmacher_survey}
   M. Mitzenmacher,
	 ``A Survey of Results for Deletion Channels and Related Synchronization Channels,''
	 \emph{Probab. Surveys}, vol.~6, pp.~1--33, 2009.
\bibitem{mundy}
   N. I. Mundy and A. J. Helbig,
   ``Origin and Evolution of Tandem Repeats in the Mitochondrial DNA Control Region of Shrikes (\emph{Lanius} spp.),''
   \emph{J. Mol. Evol.}, vol.~59, no.~2, pp.~250--257, 2004.
\bibitem{pemantle+wilson}
   R. Pemantle and M. C. Wilson,
   \emph{Analytic Combinatorics in Several Variables},
   Cambridge University Press, 2013.
\bibitem{ramezani}
   M. Ramezani and M. Ardakani,
	 ``On the Capacity of Duplication Channels,''
	 \emph{IEEE Trans. Commun.}, vol.~61, no.~3, pp.~1020--1027, 2013.
\bibitem{shannon}
   C. E. Shannon,
   ``The Zero Error Capacity of a Noisy Channel,''
   \emph{IRE Trans. Inf. Theory}, vol. 2, no. 3, pp. 8--19, 1956.
\bibitem{yeung}
   R. W. Yeung, N. Cai, S.-W. Ho, and A. B. Wagner,
	 ``Reliable Communication in the Absence of a Common Clock,''
	 \emph{IEEE Trans. Inf. Theory}, vol.~55, no.~2, pp.~700--712, 2009.
	
\end{thebibliography}
\end{document}